\documentclass[reprint, amsmath,amssymb, aps,]{revtex4-1}

\usepackage{graphicx}
\usepackage{dcolumn}
\usepackage{bm}
\usepackage{nicefrac}
\usepackage{slashed}
\usepackage{braket}

\usepackage{color}

\begin{document}

\preprint{APS?}

\title{A Novel Non-Perturbative Lattice Regularization\\
 of an Anomaly-Free $1+1d$ Chiral $SU(2)$ Gauge Theory}

\author{Michael DeMarco}
\affiliation{
Department of Physics, Massachusetts Institute of Technology, Cambridge, Massachusetts 02139, USA.
}

\author{Xiao-Gang Wen}
\affiliation{
Department of Physics, Massachusetts Institute of Technology, Cambridge, Massachusetts 02139, USA.}
\date{\today}

\begin{abstract}
We present a numerical treatment of a novel non-perturbative lattice regularization of a $1+1d$ $SU(2)$ Chiral Gauge Theory. Our approach follows recent proposals that exploit the newly discovered connection between anomalies and topological (or entangled) states to show how to create a lattice regularization of any anomaly-free chiral gauge theory. In comparison to other methods, our regularization enjoys on-site fermions and gauge action, as well as a physically transparent fermion Hilbert space. We follow the `mirror fermion' approach, in which we first create a lattice regularization of both the chiral theory and its mirror conjugate and then introduce interactions that gap out only the mirror theory. The connection between topological states and anomalies shows that such interactions exist if the chiral theory is free of all quantum anomalies. Instead of numerically intractable fermion-fermion interactions, we couple the mirror theory to a Higgs field driven into a symmetry-preserving, disordered, gapped phase.
\end{abstract}

\maketitle

Taming the ultra-violet divergences of Quantum Field Theories (QFTs) by
defining them on a lattice has been an invaluable tool in the study of both
high-energy and condensed matter physics. However, in chiral QFTs, such as the
Standard Model, the gauge fields couple to left- and right-handed modes
differently. Defining lattice chiral QFTs has presented an ongoing challenge.
Nielson and Ninomiya \cite{NIELSEN1981219} first pointed out that, for
na\"{i}ve free band theories, the periodic nature of band structure implies
that any gauge field must couple to right- and left-handed modes in the same
way. Since then, numerous approaches have tried to sidestep this no-go result.
In this paper, we present a numerical demonstration of a novel but simple
lattice regularization method that uses interactions with a disordered Higgs
field to avoid the no-go theorem while still remaining amenable to simulation.  

This work follows many other approaches to the lattice chiral QFT problem. One
of the most successful approaches to chiral Lattice gauge theory is the class
of Ginsparg-Wilson theories \cite{PhysRevD.25.2649}, which often involve non-on-site actions of the gauge symmetry. Another remarkable class of theories are the Overlap-Fermion approaches \cite{NARAYANAN199362, NARAYANAN1997360,
Luscher:1998du, Luscher:2000hn}, which compute correlation functions as the
overlaps of successive ground states. However, it is unclear if these solutions
possess finite dimensional Hilbert spaces for finite space volumes. The domain
wall \cite{KAPLAN1992342, SHAMIR199390} technique is the predecessor to what we
employ, though in that approach the gauge fields propagate in one higher
dimension than the fermions. 

The approach used in this paper, proposed previously in Refs. \cite{Wen:2013ppa,Wang:2013yta,YX14124784, PhysRevLett.116.211602}, follows the `mirror fermion' approach \cite{Mirror1, Montvay:1992eg, Giedt:2007qg, PhysRevD.94.114504}, in which one first creates a lattice regularization of both the anomaly-free chiral theory and its mirror conjugate, and then introduces interactions to gap out only the mirror theory and leave the chiral theory unchanged. The mirror fermion approach enjoys on-site fermions and gauge symmetry, as well as a physically transparent, finite dimensional Hilbert space for each site. Previous attempts at gapping out only anomaly-free mirror fermions without breaking the gauge symmetry have been unsuccessful \cite{Golterman:1992yha, Lin:1994pv, Chen:2012di, Banks:1992af}, which led to speculation that it is impossible to do so \cite{Banks:1992af}.  On the other hand, recently discovered connections \cite{Wen:2013oza, Wang:2014tia} between entangled---or topological---states and quantum anomalies suggest that we should be able to gap out only the mirror theory without breaking any gauge symmetry if the chiral theory is free of all anomalies. However, an explicit symmetric lattice model that gaps out only the mirror sector has not been constructed, and it is not known if such a model really exists. Furthermore, the gapping of the mirror sector cannot be realized by any fermion mass terms, which always break the $SU(2)$ symmetry. Since the original proposals, a number of arguments have arisen suggesting $SU(2)$ symmetric gapping mechanisms for fermions \cite{BenTov:2014eea,PhysRevD.93.081701,Ayyar2016,BenTov:2015gra}. Here we present the numerical results of a simple example $1+1d$ chiral theory with gauge group $G=SU(2)$.  For $1+1d$ chiral theories, the left- and right-handed excitations simply become spinless left- and right-moving complex fermions. In our model, the 8 left- and 8 right-moving fermions carry the following $SU(2)$ representations: $1_{R}\oplus (0_{R})^{5}\oplus (\nicefrac{1}{2}_{L})^{4}$.  We show that the symmetric lattice model that gaps out only the mirror sector indeed exists, despite the fact that $SU(2)$ symmetric quadratic mass terms cannot exist.

\begin{figure}
\includegraphics[width=.5\textwidth]{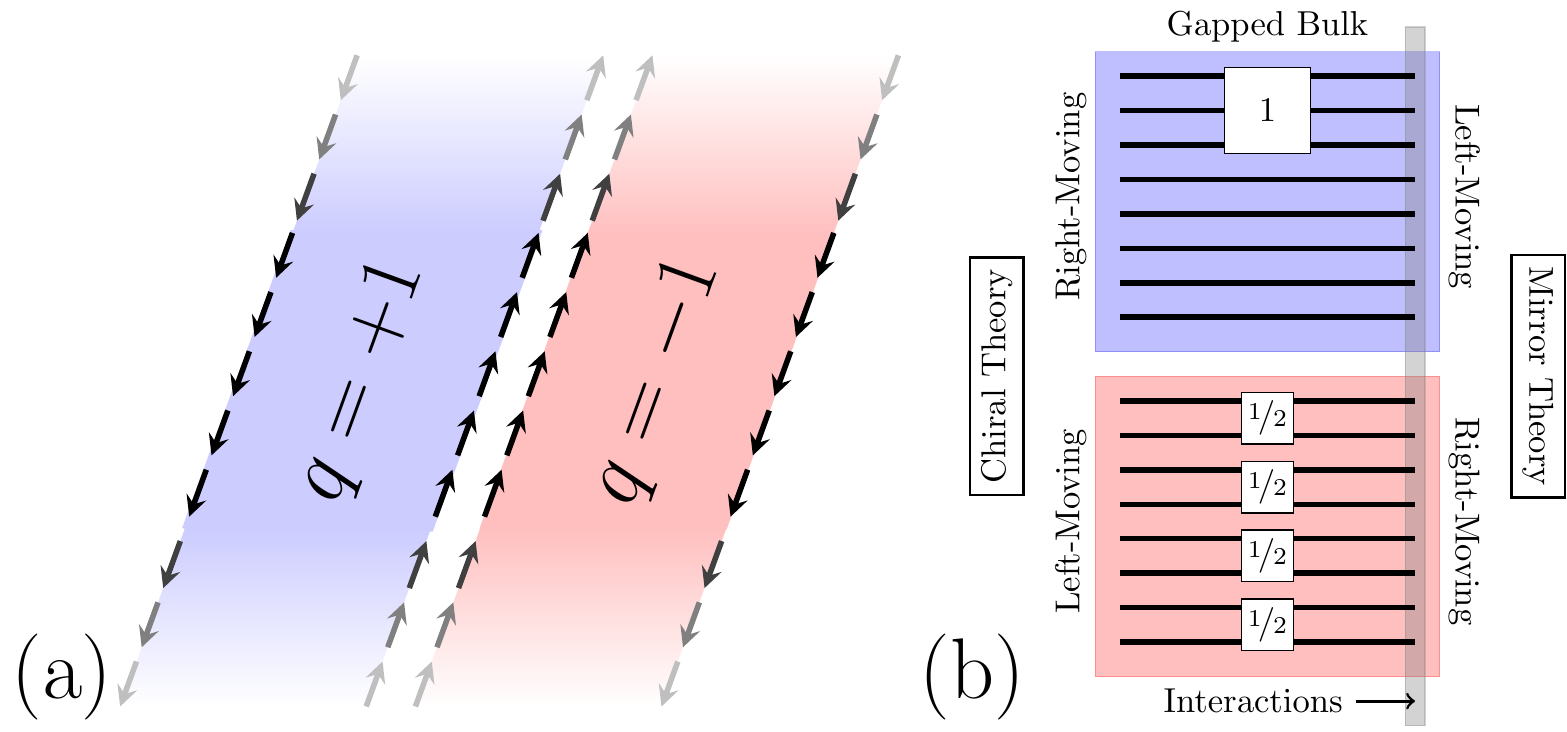}
\caption{(color online) (a) Schematic of an IQH state. Each IQH state has a single low-energy mode on each boundary but is otherwise gapped. (b) General schematic of our model. Each black line represents a $2+1$ dimensional lattice structure. Each $1+1d$ edge gives rise to a single left- or right-moving mode. We stack $8$ $\nu=1$ IQH states (blue) with $8$ $\nu=1$ parity-reversed IQH states (red). Next, we organize these modes into $SU(2)$ representations. Each representation is illustrated by a white square with the spin label except the trivial spin-0 representations which we omit. This results in a $1_{R}\oplus (0_{R})^{5}\oplus (\nicefrac{1}{2}_{L})^{4}$ theory on the left edge and its mirror $(L\leftrightarrow R)$ theory on the right edge. We couple the fermions on the right edge to a Higgs field that gaps out the right edge, leaving only the chiral theory on the left edge.}\label{fig:modelschem}
\end{figure}

Our approach begins by creating a lattice model with two edges (Figure \ref{fig:modelschem}). The system is gapped in the bulk with the only low-energy excitations localized at edges; one edge will be described by the chiral QFT and the other will be described by its mirror conjugate QFT. To do so, we construct a space-time lattice Integer Quantum Hall (IQH) state with filling fraction $\nu=1$ extending in the $t, x,$ and $w$ directions, with open boundary conditions in the $w$ direction and periodic boundary conditions otherwise (see appendix A for an explicit construction). We take $L_{t}=L_{x}\equiv L\gg L_{w}$, fixing $L_{w}$ while scaling $L$ so that the system is effectively $1+1d$. 

An IQH state gives rise to one linearly dispersing mode on each boundary. Working in Wick-rotated space, let us denote the spacetime lattice Lagrangian for an IQH $\nu=1$ state with charge $+1$ as $\slashed{D}_{L}$. Fixing a sign convention, the low energy description of $\slashed{D}_{L}$ is $\psi^{\dagger}(\partial_{t}-i\partial{x})\psi$ for $w=1$ and $ \psi^{\dagger}(\partial_{t}+i\partial{x})\psi$ for $w=L_{w}$; these are the left and right-moving modes, respectively. We can repeat this construction with charge $-1$ to get a spacetime lattice Lagrangian $\slashed{D}_{R}$. At low energies, $\slashed{D}_{R}$ has a right moving mode at $w=1$ and a left moving mode at $w=L_{w}$. We can stack IQH states to form a system of $n_{R}$ right-moving and $n_{L}$ left- moving fermions at $w=1$, subject to the fact that the $w=L_{w}$ edge realizes a system of $n_{L}$ right-moving and $n_{R}$ left-moving modes.

Next, we introduce interactions only to the mirror edge at $w=L_{w}$ that gap out all low-energy modes there. Whether or not this is possible, and how to do so, was elucidated in the recent proposals that led to this paper \cite{Wen:2013ppa,Wang:2013yta}. There is a deep connection \cite{Wen:2013oza, Kong:2014qka, Wang:2014tia} between anomalous theories and entangled, or topological, states that we briefly review here. Entangled states that have entanglement structures which are short ranged but protected by symmetry are called Short-Range Entangled (SRE) or Symmetry-Protected Topological states \cite{Chen:2010gda, Chen1604,Chen:2011bcp}; those with long-range entanglement are Long-Range Entangled (LRE) or Topological states. Remarkably, anomalous field theories that suffer gauge or gravitational anomalies can be realized as the low-energy boundary theories of SRE or LRE states, respectively. The disastrous effects of the anomalous symmetry on the edge are canceled by other edges or by a bulk which is gauge invariant only up to a surface term; the anomaly here is caused by attempting to treat the edge alone when it is non-trivially entangled with other parts of the system. Conversely, the edge theories of trivially entangled states, which possess neither protected nor long-ranged entanglement, are anomaly free, and therefore we may consider anomalous theories simply as edge theories of entangled states.  

The aforementioned connection between anomalies and entangled states has important implications for our chiral lattice regularization. Having created a lattice system with a $G$ flavor symmetry, suppose that we have successfully introduced interactions to the mirror edge at $w=L_{w}$ that, without breaking the $G$ symmetry or introducing any additional gapless modes, have gapped out the mirror edge. Entangled states generically possess gapless edge modes \footnote{There are plenty of exceptions to this, particularly when the mirror edge may be given a gapped topological order (e.g. \cite{2013PhRvX...3d1016F,Chen:2013jha}). Though such exceptions may lead to unique special cases, the generic, naive approach we present requires a trivially entangled bulk.}, and so the system must be in an unentangled (product) state near the now-gapped mirror edge. As there are no further gapless excitations in the bulk, we conclude that the entire bulk must be in a trivial, gapped phase. Following the preceding discussion, this implies that the chiral theory, which contains all remaining gapless modes, must be anomaly-free.

Confirming that a theory is free of all anomalies is not generally an easy task. The cancellation of all Adler-Bell-Jackiw (ABJ) \cite{Bell1969,PhysRev.177.2426} type anomalies can be ensured using the usual anomaly cancellation conditions \cite{Peskin:257493, Wang:2013yta}, which examine the Lie Algebra of the gauge group $G$ to provide powerful constraints. However, anomalies beyond those detectable from the Lie Algebra can still occur (e.g. \cite{WITTEN1982324}) which result from the non-trivial homotopic structure of the gauge group. For our $1+1d$ system defined on the edge of a $2+1d$ bulk, we ought to na\"{i}vely require that $\pi_{n}(g)=0$, $n\leq3$. 

For this paper, we take the Lie Group $G=SU(2)$. The simplest $SU(2)$
representation that satisfies the anomaly cancellation conditions is
$1_{R}\oplus(0_{R})^{5}\oplus(\nicefrac{1}{2}_{L})^{4}$, where subscripts
indicate a collection of left- or right-movers. Topologically, $SU(2)\simeq
S^{3}$, and so while $\pi_{1}(SU(2))=\pi_{2}(SU(2))=0$,
$\pi_{3}(SU(2))=\mathbb{Z}$. Fortunately, this simply reflects the possibility
of a Wess-Zumino-Witten (WZW) \cite{WESS197195} term, and corresponds to
the perturbative ABJ anomalies which are absent in our model by
design. 

\begin{figure*}
\includegraphics[width=\textwidth]{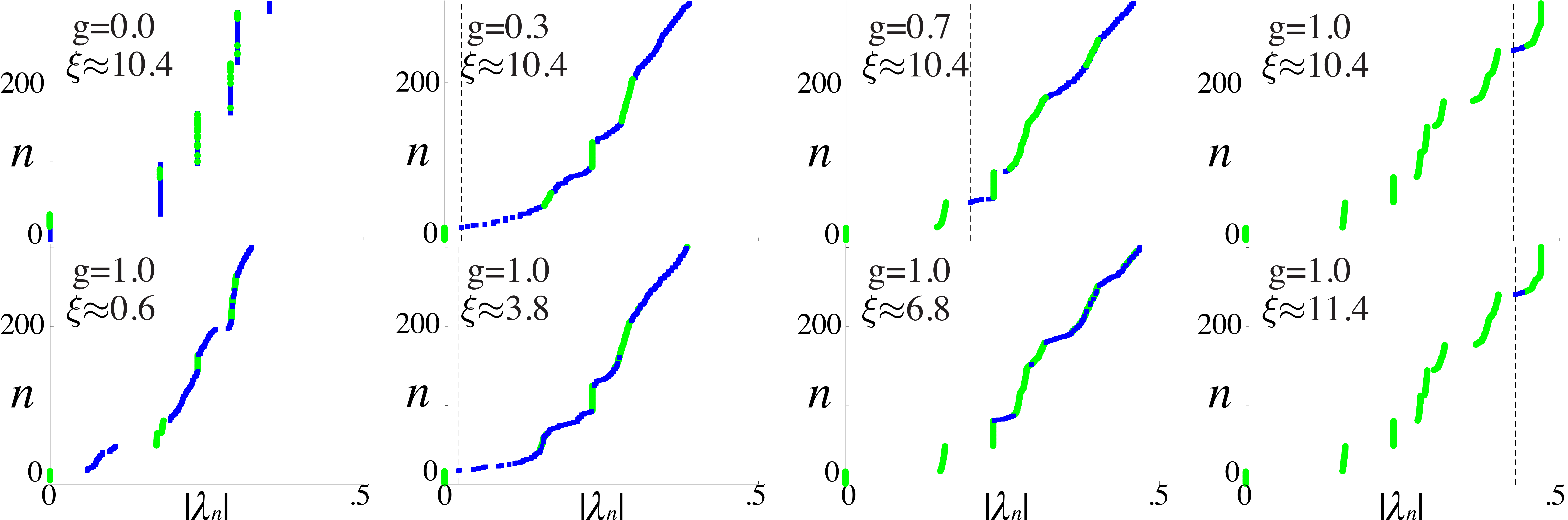}
\caption{(color online) Integrated Density of States (IDOS) for various choices of the coupling strength $g$ and the correlation length of the Higgs field $\xi$ with $L=80$. We first find the low-magnitude eigenvalues and order them in as $|\lambda_{n}|\leq |\lambda_{n+1}|$. Each plot shows $n$ vs. $|\lambda_{n}|$. States localized on the chiral edge are denoted by green circles, all others are denoted by blue squares. A black dashed line indicates the magnitude of the lowest mirror-edge eigenvalue. In the upper panels, we use the same configuration of the Higgs field with $\xi\approx 10.4$ and slowly turn on the interaction. For $g=0$, the IDOS is just that for two copies of our chiral edge theory. As the interaction strength increases, states not localized to the chiral edge are gapped out until, at $g=1$ only the chiral theory remains (below $|\lambda|\lesssim.45$). The resulting IDOS is just that for the chiral theory. The slight momentum renormalization can be mitigated by increasing $\ell_{w}$, though at high computational cost. In the lower panels, we fix $g=1$ and progressively smooth the Higgs field, increasing the Higgs correlation length $\xi$ from $\xi\approx .6$ until $\xi\approx 11.4$. For $\xi\approx .6$, there are still many low-energy mirror edge states, though their momentum structure is wiped out by the strongly disordered $\phi$. As we increase $\xi$, the mirror edge gap increases until at $\xi\approx 11.4$ no mirror-edge states remain (below $|\lambda|\lesssim.45$).}\label{fig:IDOS}
\end{figure*}

Having built a lattice theory that gives rise to a chiral theory and its mirror conjugate, we now must choose the $SU(2)$-symmetric interactions which will gap out the mirror theory. Fermion-fermion interactions would render the problem numerically intractable. Instead, we use an $SU(2)$ Higgs field. We consider the field to be condensed $|\phi(x)|=1$, leaving a non-linear $\sigma$-model. In contrast to the usual, symmetry-breaking $\phi=\text{const.}$ Higgs configurations, we drive the Higgs field into a disordered, symmetry-preserving phase with zero spatial average $L^{-2}\int d^{2}x\phi(x)=0$. We demand that $\phi$ fluctuates smoothly, with a correlation length $\xi\gg1$ that remains finite when we scale the system size. Capturing the fluctuations of this dynamical Higgs field is precisely why our calculation must be done on a spacetime lattice. 

The choice of dynamical Higgs field is of central importance. If at any point $\phi$ fluctuates too rapidly, a low-energy fermion mode may be localized there. An ideal configuration would have $|\nabla\phi(x)|=\text{const.}>0$. In the lattice model, we first choose a random $\phi(x)$ and then smooth it, taking care to apply the most smoothing in regions of largest $|\nabla\phi|$. This nonlinear smoothing process leads to a $\phi$ with nearly constant but nonzero $|\nabla \phi|$. We note that our approach resembles previous notions of condensing fluctuations \cite{You:2014vea}, as we condense $\phi$ to set $|\phi|=\text{const.}$ then condense $\nabla\phi$ to set $|\nabla\phi|=\text{const}$. 

We can now assemble the full lattice model. For our chosen $SU(2)$ representation, we need $8$ left-moving and $8$ right-moving modes, which we collect into $\Psi_{L}, \Psi_{R}$, respectively. For a given $\phi \in SU(2)$, denote by $\Theta_{L, R}[\phi]$ the block diagonal representation of $\phi$ acting on the left- and right-moving modes, respectively. Then the total Lagrangian is
\begin{equation}
\Psi^{\dagger}\slashed{D}\Psi=\Psi^{\dagger}\left(\begin{array}{cc}\slashed{D}_{R} & g\Theta_{R}^{\dagger}\Theta_{L} \\ g\Theta_{L}^{\dagger}\Theta_{R} & \slashed{D}_{L} \end{array}\right)\Psi
\end{equation} 
where $\Psi^{\dagger}=(\Psi_{R}^{\dagger}, \Psi_{L}^{\dagger})$, $\slashed{D}_{L,R}$ are the spacetime Lagrangian matrices for the individual IQH states described above, and we have introduced a coupling $g$. Here, $\Theta_{L, R}[\phi(x, t)]$ encode the $1_{R}\oplus(0_{R})^{5}\oplus(\nicefrac{1}{2}_{L})^{4}$ action of the Higgs field, though we have suppressed the Higgs dependence above. Note that $\slashed{D}$ is itself a Lagrangian, not a Langrangian density, so the fermion action is just $S[\Psi^{\dagger}, \Psi]=\Psi^{\dagger}\slashed{D}\Psi$.

The full partition function of our system is now:
\begin{equation}
\begin{aligned}
Z=\int D\phi e^{-S_{\text{H}}[\phi]}\int D\Psi^{\dagger}D\Psi\exp[-\Psi^{\dagger}\slashed{D}\Psi]\\
=\int D\phi e^{-S_{\text{H}}[\phi]}\det(\slashed{D})\label{eq:PartFun}
\end{aligned}
\end{equation}
where $S_{H}[\phi]=-U[\phi]$ is the action for the Higgs and we have neglected a proportionality constant. Performing the full integral is intractable. Instead we adopt a semi-classical approach and perform the calculation for a few $\phi$ configurations produced via the nonlinear smoothing process. These configurations provide a representative sample. Moreover, because we have chosen $S_{H}[\phi]$ to give rise to a disordered phase, any calculation is spatially self-averaging. To perform the calculation, we simply create a Higgs configuration and then assemble and calculate the small eigenvalues of the matrix $\slashed{D}$. 

Recalling that the Lagrangian density gives rise to low-energy modes of the form $\psi^{\dagger}(\partial_{t}\pm i\partial_{x})\psi$, we see that $\slashed{D}$ is not Hermitian. The eigenvalues of $\slashed{D}$, denoted $\lambda_{n}$, will generically be complex. However, we can quickly see how to interpret the eigenvalues. In momentum space for small $\omega$ and $k$, the low energy theories are of the form $\psi^{\dagger}(i\omega -H(k))\psi$, with $H(k)=\pm k$. From this we can quickly read off the meaning of the eigenvalues at low frequency $\omega\ll 1$: the real part of the eigenvalue corresponds to the energy, while the imaginary part is the variation in time. Hence any gapless excitations can be identified by their $\omega\to 0$ limits. In turn, a gap is just a region around zero in the complex plane devoid of eigenvalues, and we take $\Delta=\min\{|\lambda_{n}|\}$. 

We are now ready to perform the numerical calculation. To see what the process of gapping out the mirror theory looks like, let us first fix the system size $L=80$ and Higgs correlation length $\xi$. Figure \ref{fig:IDOS} (upper panels) show the integrated density of states as we turn on the interaction from $g=0$ to $g=1$. At $g=0$, there are 32 gapless modes---16 from the chiral theory and 16 from the mirror conjugate. As we turn on the interaction, the fluctuating Higgs field smoothly gaps out the mirror theory modes, leaving only the chiral theory at low energies. 

Next, we fix $g=1$ and instead vary the Higgs field correlation length $\xi$ in Figure \ref{fig:IDOS} (lower panels). For $\xi \lesssim 2$, the mirror edge modes appear at small magnitude, though their momentum structure is wiped out by the rapidly fluctuating Higgs field. As $\xi$ is increased, the mirror theory modes again are driven to large magnitude, leaving only the chiral theory at small eigenvalues. 

\begin{figure}
\centering
\includegraphics[width=.45\textwidth]{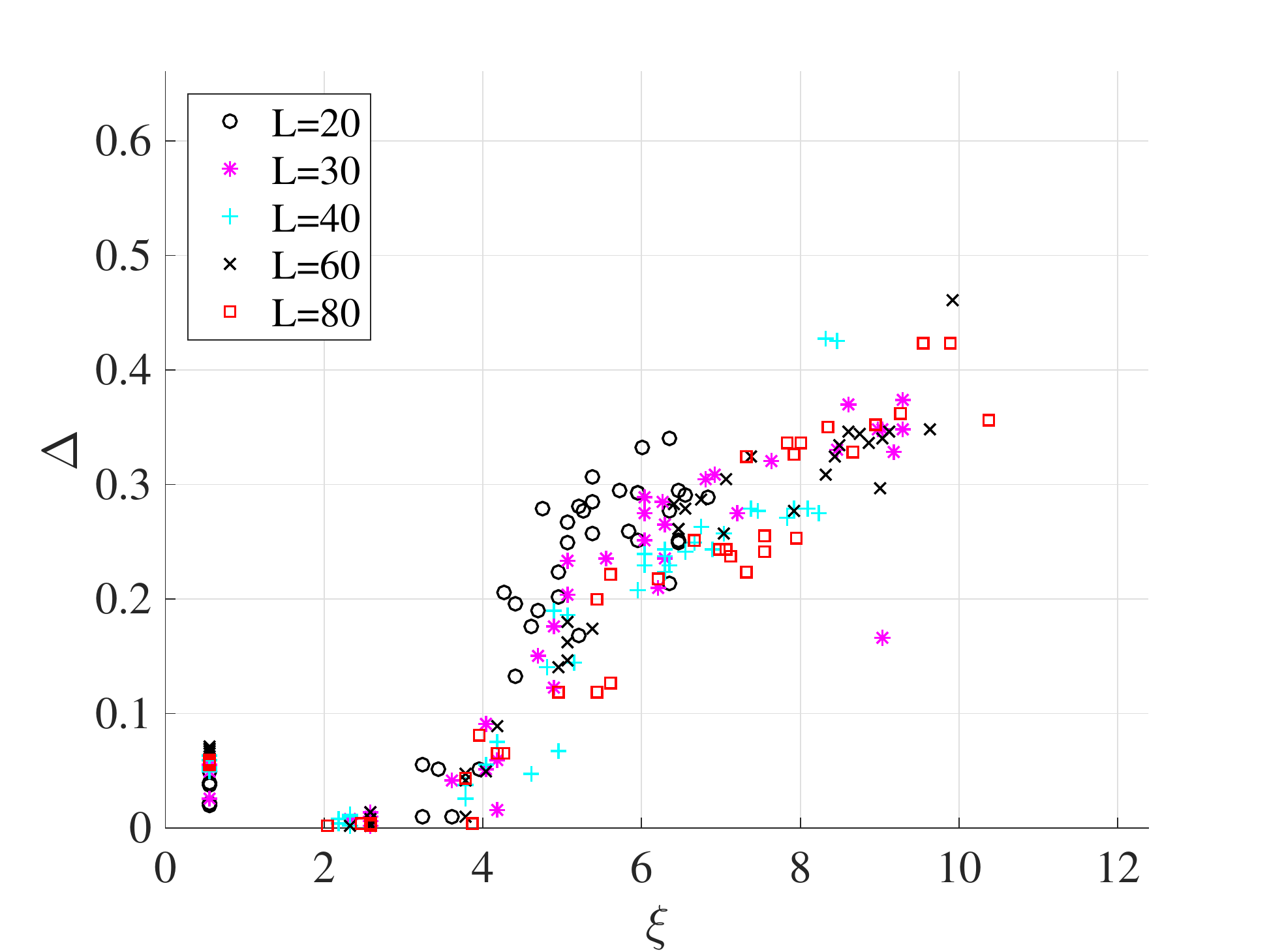}
\caption{(a) Mirror theory gap as a function of correlation length $\xi$ for various choices of $L$. We couple the fermions on the mirror theory edge to a Higgs field fluctuating with correlation length $\xi$ and look for the smallest magnitude eigenvalue for states not strongly localized on the chiral theory edge. $\xi=L$ would correspond to the usual symmetry-breaking Higgs mechanism. We choose a finite $\xi$, independent of $L$. For $\xi \gtrsim 8$, the mirror edge is gapped with $\Delta\approx .35$, and we see that this gap is independent of $L$ for $L\gtrsim 40$, indicating thermodynamic behavior. Note that, in the $\xi=L\to\infty$ `symmetry-breaking' limit, $\Delta=g=1$.}
\label{fig:thermo}
\end{figure}

Now we turn to the thermodynamic behavior, fixing $g=1$ and examining the gap given to the mirror modes. Figure \ref{fig:thermo} shows how this mirror-mode gap scales with the system size $L$ and the correlation length $\xi$. At $\xi\approx 8$, for system sizes $L\gtrsim40$, we see that the gap is roughly constant at $\Delta\approx.35$, and largely independent of system size. This indicates that we are indeed seeing thermodynamic-limit behavior and that we have successfully gapped out only the mirror theory. A related test of thermodynamic behavior is demonstrated in Appendix \ref{sec:LargerLattice}.

While we have seen that there is a spectral gap $(\det(\slashed{D})\neq 0)$ in the fermionic sector of the mirror theory, we must now discuss why the partition function is not wiped out by the integral over Higgs configurations. Since $\slashed{D}$ is not Hermitian, this determinant may be complex and its phase can fluctuate. Fortunately, anomalous behavior is prevented by the anomaly-free conditions discussed earlier: first we note that $\pi_{i}(SU(2))=0$, $i=0, 1, 2$, so there are no topologically distinct sectors of Higgs field configurations or topological defects; secondly, the presence of a perturbative WZW term is prevented by the anomaly cancellation condition. 

There is a symmetry that controls the phase of the partition function. Consider first the time reflection symmetry that sends $\phi(x, t)\to\phi(x, -t)$ or, equivalently, reverses the lattice in the time direction and leaves $\phi(x, t)$ unchanged. Because the imaginary parts of eigenvalues come only from the temporal hopping, this has the effect of complex conjugation $\lambda_{n}\to \lambda_{n}^{*}$. This implies that so long as the action for $\phi$ is time-reflection symmetric, we may replace the determinant by its real part in the partition function:
\begin{equation}
Z=\int D\phi e^{-S[\phi]}\Re[\det(\slashed{D})]
\end{equation}
Hence the partition function is real, despite involving a manifestly non-Hermitian Lagrangian.

Finally, we wish to highlight the difficulty of writing a field-theoretic
description for the mirror edge in this model. The na\"{i}ve approach of
employing a theory of Weyl fermions and a Higgs field is spoiled by the Higgs
potential: a Higgs configuration with $|\nabla\phi|=\text{const.}>0$ can be
realized as minimizing the potential $U[\phi]=\int dx dt[-a|\nabla
\phi|^{2}+b|\nabla\phi|^{4}+...$ with $a,b>0$. However, we do not know at
present how to evaluate such a field integral. Any more elegant attempt is likely to
be complicated, as there is no mass term capable of giving a mass to the
fermion fields without breaking symmetry. Whether a low-energy description can
be found in terms of fields radically different than the na\"{i}ve fermion
fields---or in terms of a CFT without a Lagrangian description---remains to be
seen.

In summary, the method that we have demonstrated in this paper leads to a particularly simple lattice regularization for chiral QFTs. Both fermions and the gauge symmetry action are entirely onsite, and the Hilbert space associated to any one site is well-defined and physically clear. We consider this model's inability to regularize most anomalous theories a great feature; this method brings the physical nature of quantum anomalies nearer to heuristic focus. Crucially, the use of a Higgs field, as opposed to fermion-fermion interactions, renders this method very promising for further numerical studies.  

This material is based upon work supported by NSF Grant No. 1122374,  Grant No.
DMR-1506475, and NSFC 11274192.  M.D. acknowledges revisions from and discussions with H. Pakatchi, M. Pettee, and L. Liu.

\bibliography{BibBank}

\appendix
\section{Lattice Model}
Here we give the explicit construction of the spacetime Lagrangian. We first build a spatial, lattice model of a single IQH state before extending the model to give a spacetime structure. Then, we stack copies of these spacetime IQH models to give our full 16-flavor model. 

\begin{figure}
\centering
\includegraphics[width=.5\textwidth]{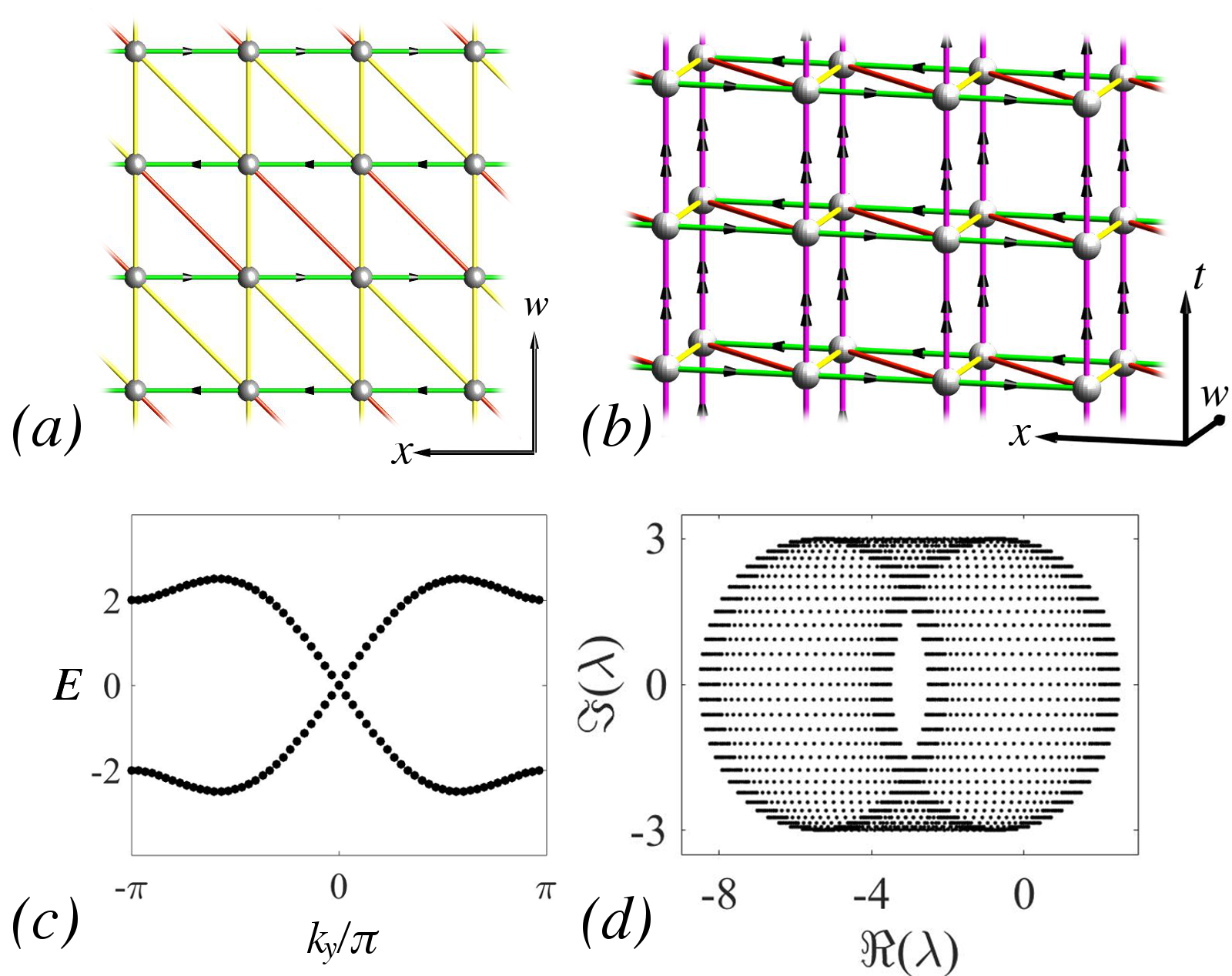}
\caption{(color online) (a): Spatial IQH state hopping model. Fermion sites are shown as spheres, with hopping terms as links. A yellow link indicates a hopping of $+1$, a red link a hopping of $-1$, and a green link a hopping of $+i$ in the direction of the arrow and $-i$ in the opposite direction. Hopping around any plaquette generates a phase of $\pi$, hence with $1$ fermion per site this is a $\nu=1$ IQH state. (b): Spacetime lattice with $L_{w}=2$. Each spatial component is just a slice of a IQH state shown in (a), while the purple links represent a Hopping of $ti$ in \emph{only} the direction of the double arrows. In addition, each site is given the onsite chemical potential $-t\psi^{\dagger}_{i}\psi_{i}$, where we later set $t=3$.The Hopping matrix corresponding to this model is precisely our spacetime Lagrangian. (c) Dispersion relation for the spatial lattice with $L_{w}=2$. (d) Complex Eigenvalues of the spacetime hopping model. }\label{fig:LatticeDetail}
\end{figure}

The spatial component of our lattice is provided by the lattice Integer Quantum Hall (IQH) states shown in figure \ref{fig:LatticeDetail} (a) extending in the $w-x$ plane. Note that hopping around any plaquette gives a phase of $\pi$, hence we have exactly one flux quantum per site. This model is tuned so that the chemical potential is exactly between the Landau levels, giving rise to a $\nu=1$ state. We give the $x$ direction periodic boundary conditions but leave open boundaries in the $w$ direction. The open boundary conditions give rise to two low-energy modes: a right-moving excitation on the right hand edge and a left-moving excitation on the left hand edge. Denoting this matrix as $H_{R}$, the $q=-1$ state is just obtained by $H_{L}=H_{R}*$.

In our calculation, we assumed that the two edges were effectively decoupled. Na\"{i}vely, this would require $L_{w}\gg1$, but for eigenvalue computations $L_{w}=2$ is in fact sufficient. With $L_{w}=2$, momentum spectrum of the chiral theory is slightly affected by the Higgs fluctuations on the mirror edge. This effect may be reduced by increasing $L_{w}$, at significant computational cost. The eigenvalues of the $L_{w}=2$ system are shown in figure \ref{fig:LatticeDetail}c; the twin linear modes are precisely the edge modes.

We can then add a time hopping to our spatial system to create a spacetime lattice as shown in figure \ref{fig:LatticeDetail}b. In momentum space this hopping leads to a term $T\equiv t(e^{i\omega}-1)$. We set the strength of the time hopping term $t$ so that near $\omega=\pi$, the real part of the eigenvalue is more negative than the bandgap of the system (note that since $\mathcal{L}=p\dot{q}-\mathcal{H}$, a negative real part is a positive energy). For our purposes, $t=3$ suffices. The one-flavor spacetime Lagrangian is then just $\slashed{D}_{L,R}\equiv T\oplus (-H_{L, R})$, where $\oplus$ is the usual Kronecker sum. 

The complex eigenvalue spectrum of $\slashed{D}_{R}$ is shown in Figure \ref{fig:LatticeDetail}d. For any fixed $\omega$, the corresponding eigenvalues form a line parallel to the real axis. The broader, circular complex structure is then produced by the $e^{i\omega}-1$ terms. 

For our full model, we need $8$ copies each of $\slashed{D}_{R}$ and $\slashed{D}_{L}$, together with the interaction discussed in the main body of the paper. The total form is then:
\begin{equation}
\Psi^{\dagger}\slashed{D}\Psi=\Psi^{\dagger}\left(\begin{array}{cc}\mathbb{I}_{8}\otimes\slashed{D}_{R} & g\Theta_{R}^{\dagger}\Theta_{L} \\ g\Theta_{L}^{\dagger}\Theta_{R} & \mathbb{I}_{8}\otimes\slashed{D}_{L} \end{array}\right)\Psi
\end{equation}
where $\Theta_{L, R}[\phi(x)]$ are the block-diagonal matrices corresponding to the $SU(2)$ representation $1_{R}\oplus(0_{R})^{5}\oplus(\nicefrac{1}{2}_{L})^{4}$.

\section{Larger Lattice Calculation}\label{sec:LargerLattice}
\begin{figure}
\centering
\includegraphics[width=.45\textwidth]{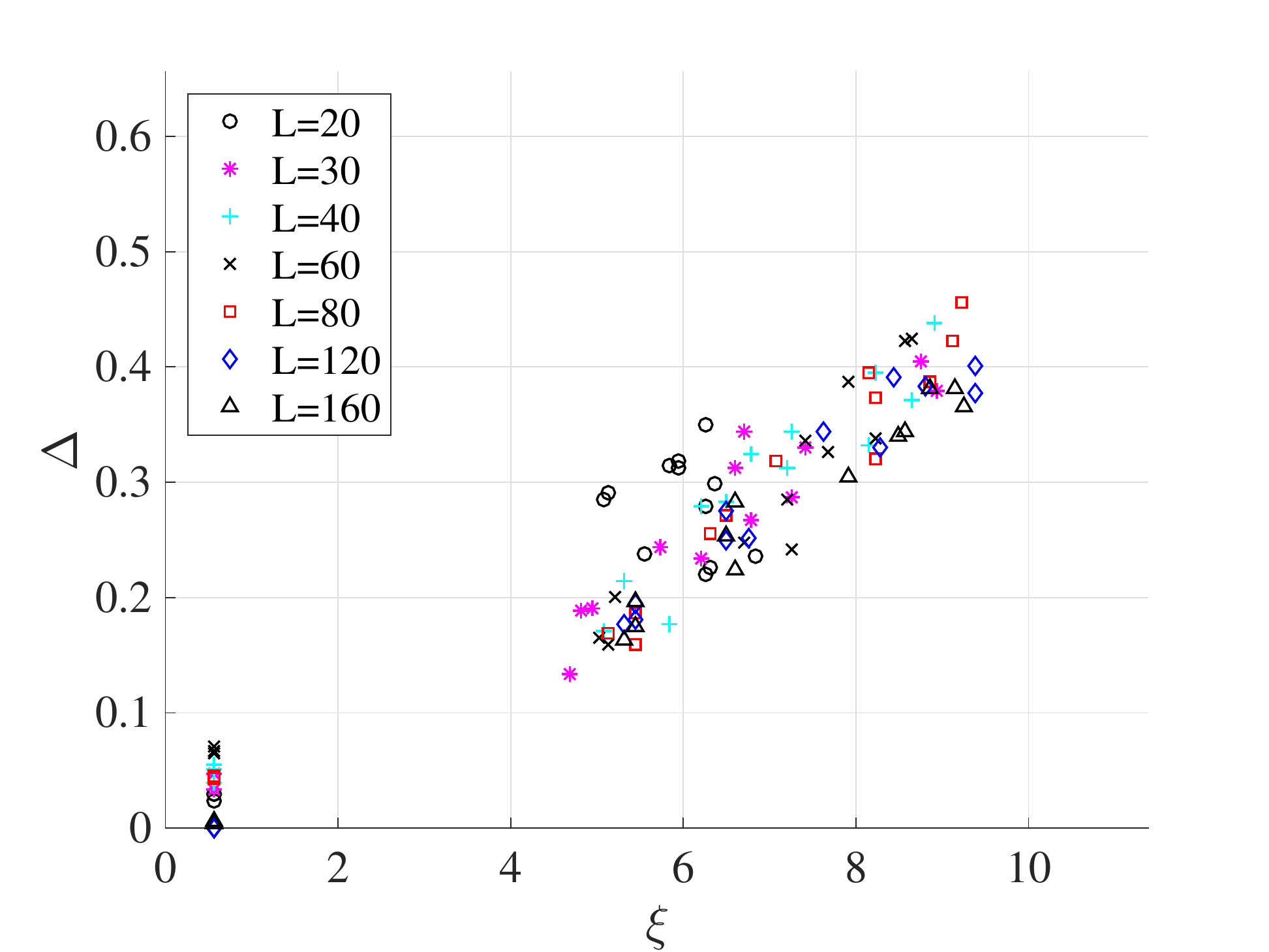}
\caption{Total gap as a function of correlation length $\xi$ for various choices of $L$. We introduce a second, independently fluctuating Higgs field to the chiral theory edge, in addition to the fluctuating Higgs field introduced on the Mirror Theory edge ($\xi$ is the minimum correlation length of the two fields). We then measure the total gap, and plot it as a function of the minimum correlation length over the two edges. This calculation allows us to extend our analysis to larger $L$ and confirm the previous results. For $\xi\gtrsim8$, the system is again gapped with $\Delta\approx .35$, and we reach thermodynamic, $L$ independent behavior for $L\gtrsim40$. Note that, in the $\xi=L\to\infty$ `symmetry-breaking' limit, $\Delta=g=1$.}\label{fig:thermo2}
\end{figure}

The main obstacle to increasing system size is the large number of chiral-edge gapless modes that must be found before the finding the first gapped, mirror-edge mode. To test the gapping effect of the disordered Higgs field for larger system sizes we introduce a second, \emph{independently} fluctuating Higgs field to the chiral theory edge and then measure the total gap. As the two edges are effectively decoupled, this encodes the same physics while allowing us to push the system size to $L=160$ with modest computational resources. Figure \ref{fig:thermo2} shows these results and demonstrates that with $\xi\approx 8$, $\Delta\approx .35$, and we are again well in the thermodynamic regime for $L \gtrsim 40$. 

\end{document}